\documentclass[aps,prl,twocolumn,superscriptaddress]{revtex4-1}
\usepackage{graphicx}

\begin{document}

\title{Intermittency and Local Heating in the Solar Wind}

\author{K.T. Osman}
\email{kto@udel.edu}
\affiliation{Centre for Fusion, Space and Astrophysics; University of Warwick, Coventry, CV4 7AL, United Kingdom}
\affiliation{Bartol Research Institute, Department of Physics and Astronomy, University of Delaware, Delaware 19716, USA}

\author{W.H. Matthaeus}
\author{M. Wan}
\author{F. Rappazzo}
\affiliation{Bartol Research Institute, Department of Physics and Astronomy, University of Delaware, Delaware 19716, USA}

\date{\today}

\begin{abstract}

Evidence for inhomogeneous heating in the interplanetary plasma near current sheets dynamically generated by magnetohydrodynamic (MHD) turbulence is obtained using measurements from the ACE spacecraft. These coherent structures only constitute $19\%$ of the data, but contribute $50\%$ of the total plasma internal energy. Intermittent heating manifests as elevations in proton temperature near current sheets, resulting in regional heating and temperature enhancements extending over several hours. The number density of non-Gaussian structures is found to be proportional to the mean proton temperature and solar wind speed. These results suggest magnetofluid turbulence drives intermittent dissipation through a hierarchy of coherent structures, which collectively could be a significant source of coronal and solar wind heating.  

\end{abstract}


\maketitle

The energy required to explain the coronal heating problem and, hence, the existence of a solar wind \citep{AxfordMcKenzie97,HabbalEA95} might be provided by magnetohydrodynamic (MHD) turbulence 
\citep{TuMarsch95,BreechEA08} driving a cascade \citep{OsmanEA11b} to scales where kinetic dissipation is efficient. This could account for the non-adiabatic expansion of heliospheric protons \citep{RichardsonEA95}. Analysis of higher order statistics from observations and numerical simulations indicates inertial range intermittency \citep{MarschTu94,HorburyEA98} is associated with coherent structures such as current sheets \citep{MattMont80, CarboneEA90,Veltri99}. These may play a role in heating the solar corona by magnetic reconnection (or nano flares). Connections between the turbulence cascade and intermittent dissipation are fundamental in hydrodynamics \citep{SreenivasanAntonia97}, but are almost unexplored in space plasma physics, which relies heavily on linear Vlasov theory under the assumption of a uniform equilibrium plasma \citep{Barnes79a,SahraouiEA09,HowesEA08}. Recently, a statistical link was found between coherent magnetic field structures and elevated temperatures \citep{OsmanEA11} (also see \citep{Burlaga68}), the interpretation of which has been questioned \citep{BorovskyDenton11}. Here evidence is presented, based on correlations at two ranges of spatial scales and on global conditional averages, that temperature enhancements are {\it locally} linked to coherent structures and intermittency in solar wind turbulence. 

Motivation for this investigation lies at the core of turbulence theory. The Kolmogorov 1941 framework \citep{Kol41a} provides a useful description, but a more precise theory must account for large fluctuations in the energy dissipation rate \citep{Obukhov62b}. The resulting intermittency can be defined as the increasing non-Gaussian character of increments with decreasing scale \citep{SreenivasanAntonia97}. It is this connection between non-uniform dissipation and the statistics of increments that is embodied in the Kolmogorov Refined Similarity Hypothesis (KRSH) \citep{Kol62}. Since solar wind turbulence is well described by ideas that parallel its hydrodynamic antecedents --- second order \citep{MattGold82a}, third order \citep{PolitanoPouquet95,PolitanoPouquet98-grl}, and higher order statistics \citep{MarschTu94} --- it is perhaps paradoxical that interplanetary dissipation is not {\it presumed} to be highly spatially non-uniform. Indeed, we are not aware of an alternative rationale that provides a physical explanation for solar wind intermittency.

A major difficulty in linking energy dissipation to intermittency within low collisionality plasmas is that the underlying dissipation mechanisms are not unambiguously known. Often solar wind dissipation is framed in terms of Landau damping, cyclotron resonance, and instabilities in a uniform plasma where linear Vlasov theory is relevant. However, the observed consistency with intermittency theory implies deeper connections to KRSH. In order to pursue an unbiased assessment, without making assumptions about the dissipation function, proton temperature is adopted as a surrogate for the dissipation. Therefore, if the dissipation is non-uniform or intermittent, it will be reflected in the temperature distribution. In particular, we would expect to find sources of temperature enhancements within and around current sheets, which are the characteristic small scale coherent structures in MHD turbulence \citep{MattMont80, MattLamkin86, CarboneEA90, Veltri99}. This is despite the collisional MHD framework not being applicable at the scales where the most intense current sheets would form. This lack of clarity regarding the balance between fluid and kinetic phenomena is partly responsible for the wide disparity of viewpoints concerning coronal and solar wind dissipation. The use of proton temperature as a surrogate side-steps this issue and addresses the statistics of dissipation without detailed knowledge of the mechanism.

We analyze the entire $64$ s resolution magnetic field and proton temperature datasets from the MAG \citep{Smith1998} and SWEPAM \citep{McComas1998} instruments onboard the ACE spacecraft at 1 AU. Rapid changes in the magnetic field are described by vector increments: 
\begin{equation}
\Delta \mathbf{b}(t,\Delta t) = \mathbf{b}(t + \Delta t) - \mathbf{b}(t)
\end{equation}
where $\mathbf{b}(t)$ is the magnetic field time series and $\Delta t$ is the time lag. Here the temperature data cadence defines the lag $\Delta t = 64$ s employed. Using Taylor's hypothesis this corresponds to a spatial separation within the inertial range, which extends from about an hour to a second in the spacecraft frame \cite{TuMarsch95}. In order to identify coherent (non-Gaussian) structures in the solar wind, a time series of the normalized partial variance of increments (PVI, denoted I in Fraktur script) is constructed:
\begin{equation}
\Im = \frac{\left| \Delta\mathbf{b} \right|}{\sqrt{\langle \left| \Delta\mathbf{b} \right|^{2} \rangle}}
\end{equation}
where $\langle \cdots \rangle$ denotes an ensemble average. Events are selected by imposing thresholds on PVI, leading to a methodology that is comparable to classic magnetic discontinuity identification in both solar wind observations and numerical simulations \citep{GrecoEA09}.

\begin{figure}[h]
\includegraphics[width=8.3cm]{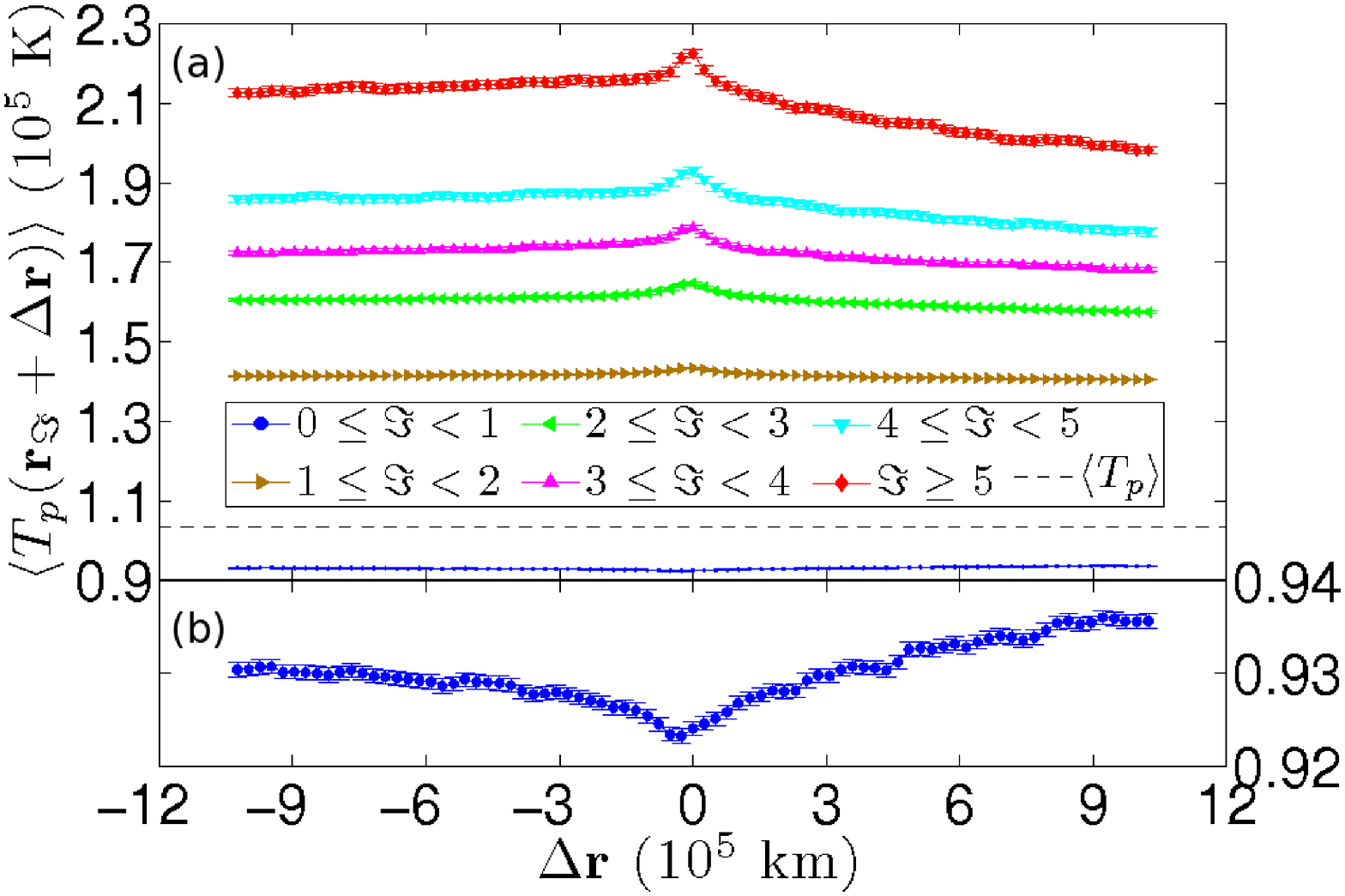}
\caption{Mean $T_{p}$ conditioned on the distance from (a) events identified using the PVI statistic. These events are linked to elevated temperatures. The temperature profile associated with $0 \leq \Im < 1$ is reproduced (b) to highlight its local minimum. These low value fluctuations are heated upon encountering current sheets. For reference, the dashed line represents the mean $T_{p}$ for the entire dataset.}
\label{fig:1}
\end{figure}

In order to study heating near discontinuities, we compute averages of $T_{p}$ conditioned on the distance from PVI events which exceed a threshold value $\theta$:
\begin{equation}
\bar{T}_{p} (\Delta r,\theta) \equiv \langle T_p(r_{\Im} + \Delta r) | \theta \leq \Im(r_{\Im}) < \theta + 1 \rangle
\label{eq:cond}
\end{equation} 
where $\mathbf{r}_{\Im}$ is the position of a PVI event and $\Delta\mathbf{r}$ is the spatial lag measured relative to $\mathbf{r}_{\Im}$, obtained assuming frozen-in flow such that $\Delta\mathbf{r} \approx -\mathbf{v}_{sw}\Delta t$. 

Figure 1 shows the conditionally averaged proton temperature in the vicinity of a PVI event for selected values of $\theta$. For thresholds above and including $\theta=2$, there is a peak in the average $T_{p}$ at the location of discontinuities, as established in \citep{OsmanEA11}. The steep fall-off at nearby spatial lags produces a defined local maximum with a width of about $10^{5}$ km, which is equivalent to around a tenth of the turbulence correlation scale $\lambda_{c}$ \citep{MatthaeusEA05}. This local $T_{p}$ maximum is more pronounced near stronger discontinuities, as seen when the PVI threshold is raised. The averaged $T_{p}$ then transitions to a gently declining plateau, which is distinct for each $\theta$ value since the strongest PVI events are associated with the highest temperatures. Analysis of the data suggests that once a strong discontinuity is identified, there is an elevated probability of finding additional strong events within several correlation lengths (of order $10^{6}$ km). The net effect of this clustering, or non-Poisson property \citep[see][for details]{GrecoEA09-wait}, of strong PVI events is that the surrounding plasma is, on average, hotter.  	 

The cumulative mean waiting distances between non-Gaussian events are consistent with the interpretation of clustering PVI events. Beyond a spatial separation of around $0.1\lambda_{c}$, there is an increased likelihood of encountering another discontinuity. Therefore, we interpret the central peak of $T_{p}$ enhancement in Fig. 1 as the result of local heating by individual coherent structures, while the broader plateau of elevated temperature is the result of an increased nearby density of strong coherent structures. This clustering of heating events is indicative of a correlated inertial range intermittency process, in contrast to an uncorrelated Poisson heating mechanism. 

\begin{figure}[b]
\includegraphics[width=8.3cm] {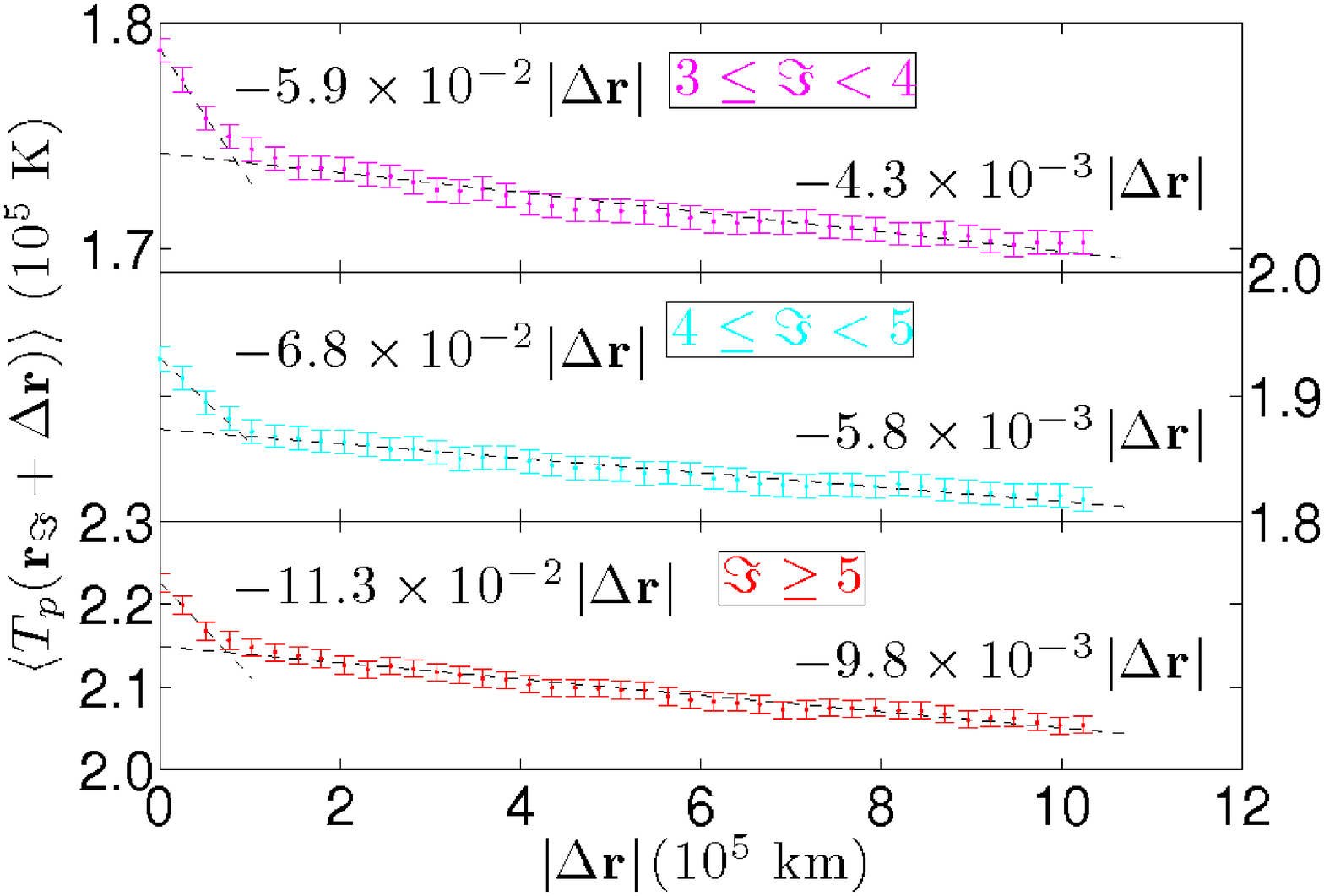}
\caption{Mean $T_{p}$ conditioned on the distance from events identified using the PVI statistic. The plots consist of a core component of enhanced temperature and an extended heated outer component. These are both well approximated by linear scaling and the obtained temperature gradients are steepest near the strongest (highest $\theta$) coherent structures.}
\label{fig:linfit}
\end{figure} 

It should be noted that all the conditional average $T_{p}$ profiles with $\Im \geq 1$ lie well above the unconditioned average value. Hence, it follows that the profile corresponding to low value fluctuations (in the range $0 \leq \Im < 1$) must lie below the average $T_{p}$ value, as shown in Fig. 1. However, it is not obvious that these fluctuations should produce a central local minimum in the conditional average temperature profile. These low level fluctuations are the smoothest regions and represent the closest approximation to uniform plasma conditions \citep{GrecoEA09}. Since the most uniform samples are cooler than the surrounding plasma, our results suggest the dominant sources of turbulence heating are unlikely to be found using methods that assume a uniform plasma.  

Figure 2 shows the conditionally averaged $T_{p}$ dependence on the magnitude of spatial separation from the central discontinuity $\left| \Delta r \right|$, for three PVI thresholds. This expanded scale allows the two-tiered structure of the averaged temperatures to be more manifest. The data is well approximated by linear fits, and the obtained temperature gradients confirm the presence of an enhanced $T_{p}$ inner core 
and an extended ($> 0.1\lambda_{c}$) heated region.   

In order to study the macroscopic behavior of intermittent heating, the data is divided into 10 hour samples and in each sample we compute the fraction of data represented by strong events satisfying $\Im \geq 4$. Figure 3(a) shows PDFs of $T_{p}$ conditioned on this density of coherent structures. As the percentage of strong PVI events increases, the probability density increases for the highest average $T_{p}$ and vanishes for the lowest. This is consistent with coherent structures being directly associated with enhancements in temperature. The 10 hour samples are sorted by average solar wind speed into bins of width 100 kms$^{-1}$, and the average coherent structure density in each bin is plotted as a function of the corresponding average $T_{p}$. Figure 3(b) shows the higher average $T_{p}$ samples also contain a greater density of strong coherent structures. A remarkable feature is that the data is ordered automatically into a sequence of increasing solar wind speed $v_{sw}$. Faster wind has higher $T_{p}$ and a higher density of discontinuities \citep{BorovskyDenton11}. The bin corresponding to the lowest wind speeds ($v_{sw} < 300$ kms$^{-1}$) lies on the same trend line, but has a lower than average $T_{p}$. The implied mutual correlation between $T_{p}$, $v_{sw}$, and heating rate is completely consistent with a turbulence-heated corona and solar wind \citep[e.g.][]{Demoulin09}. 
  
\begin{figure}[h]
\includegraphics[width=8.3cm] {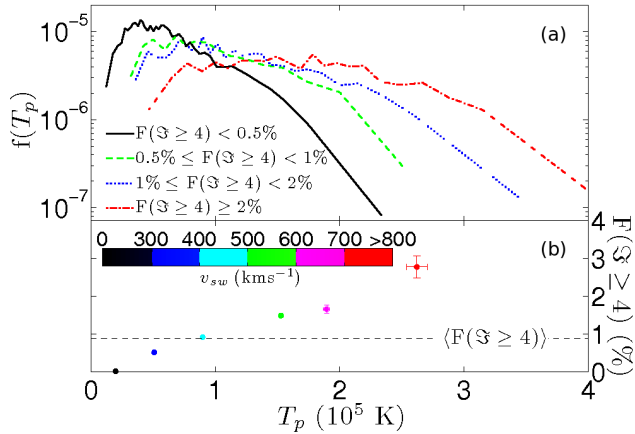}
\caption{(a) PDFs of $T_{p}$ conditioned on the density of coherent structures with $\Im \geq 4$. As the percentage of strong PVI events increases, the probability density increases for the highest average $T_{p}$ and vanishes for the lowest. (b) The higher average $T_{p}$ samples contain a higher density of strong PVI events, and are ordered according to increasing average solar wind speed.}
\label{fig:speedcorr}
\end{figure}

The evidence presented here suggests solar wind plasma is hotter in the immediate vicinity of discontinuities. The sharpest discontinuities are linked to the steepest temperature gradients and the clustering of discontinuities leads to regionally elevated temperatures. These conclusions contrast those of \cite{BorovskyDenton11} who employed a different methodology, and only studied the nearest points neighboring the local maximum. Our results suggest relationships involving the dissipation $\epsilon_{s}$ averaged over a scale $s$, the increments (on the same scale) of the MHD Els\"asser fields $\delta z_{s}$, and the local temperature $T_{p}$. Symbolically we consider two relationships:	
\begin{equation}
\epsilon_{s} \sim \frac{\delta z_{s}^{3}}{s}  \Longleftrightarrow T_{p}
\label{eq:KRSH}
\end{equation}
The first is an adaptation of the Kolmogorov Refined Similarity Hypothesis to MHD \citep[e.g.][]{MerrifieldEA05}, where $\sim$ refers to statistical distributions that result in proportionality of moments after appropriate averaging. From KRSH, which is widely supported in hydrodynamics but unproven \citep{SreenivasanAntonia97}, emerge properties such as multi-fractal scaling due to the irregular spatial distribution of $\epsilon_{s}$. In astrophysical plasmas there is generally no direct measure of $\epsilon_{s}$, although the present study does demonstrate a statistical connection between increments and local temperature enhancements: the second relationship in Eq.\ref{eq:KRSH}. Given the plausibility of a link between the local statistics of $T_{p}$ and $\epsilon_{s}$, our results support the hypothesis that MHD turbulence drives intermittent dissipation in the solar wind through kinetic processes operating within a hierarchy of intermittent coherent structures.

It is instructive to determine if the observed inhomogeneous elevations in $T_{p}$ could signal an underlying heat source that contributes significantly to solar wind heating. An initial assessment is provided by estimating the contribution from PVI events to the total internal energy $U \propto n_{p}T_{p}$, where $n_{p}$ is proton number density. Here the PVI events are conditioned on threshold $\Im > \theta$. The internal energy $U(\theta)$ and volume $V(\theta)$ are associated with the central PVI event and the two nearest and next-nearest neighbors, which together form a local maximum (see Fig. 1). Figure 4 plots the conditional internal energy density $U_{D}(\theta) = U(\theta)/V(\theta)$ normalized to the internal energy density of the entire dataset $U_{D}(\theta = 0)$. As the PVI threshold is raised, $U_{D}(\theta)$ increases as is consistent with internal energy being concentrated in strong discontinuities. In order to better quantify the contribution of coherent structures to $U_{D}$, we define non-Gaussian events as those where the ratio of PVI to Gaussian distributions exceeds an order of magnitude. For these events, the PVI threshold dependence of $U$ and $V$ is examined. Figure 4 shows events where $\Im > 2.4$ make up $50\%$ of the total internal energy but only occupy $19\%$ of the volume. The strongest PVI events, $\Im > 5$, occupy just 2\% of the volume but contribute 11\% of the internal energy. In a separate study \citep{ServidioEA11}, the threshold $\theta = 6$ was found to identify events that are likely strong active magnetic reconnection sites. Therefore, it seems the inhomogeneous features identified here could contribute substantially to the internal energy budget of the solar wind.   
 
\begin{figure}[t]
\includegraphics[width=8.3cm] {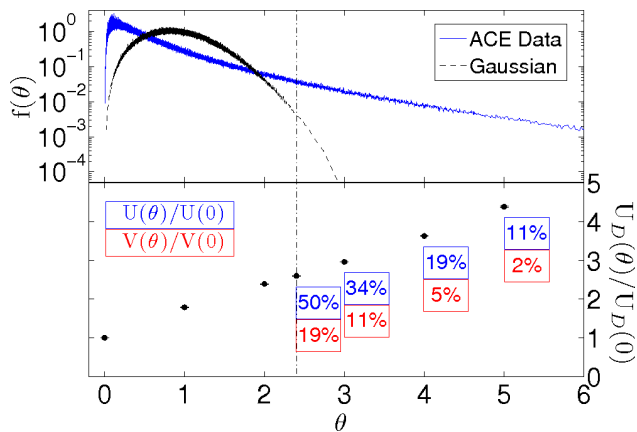}
\caption{(a) PDFs of PVI statistic from solar wind dataset compared to a unit-variance Gaussian distribution, where above $\theta = 2.4$ (dot-dash line) corresponds to non-Gaussian coherent structures. (b) Estimates of the fraction of total internal energy density $U_{D} = U/V$ in structures with $\Im > \theta$ are found to increase with $\theta$. This suggests internal energy is concentrated in discontinuities.}
\label{fig:energy}
\end{figure}

If the dissipation function is intermittent, as our results suggest, its properties can be better understood in the context of large scale turbulence, which may provide the framework to understand coronal and solar wind heating \citep{BreechEA08,CranmerEA09,VerdiniEA10}. Hence, variations in intermittent dissipation might account for differences in wind speed \citep{Demoulin09}. If demonstrated, this would bring closure to the inter-relationships between speed, temperature, and density of coherent structures, that remain otherwise presently unexplained. As observations probe below the ion inertial scale to study the kinetic physics of dissipation \citep{LeamonEA99,SahraouiEA09} and closer to the solar wind sources in the Solar Probe and Solar Orbiter missions, the link between inertial range increments and non-uniformity in dissipative structures will become more important. Indeed, such efforts might establish a much closer relationship between solar wind turbulence and the classical hydrodynamic concept of intermittent dissipation \citep{Kol62,Obukhov62b}.

This work was funded by STFC, the NSF SHINE (ATM-0752135) and Solar Terrestrial (AGS-1063439) Programs, and by NASA under the Heliophysics Theory (NNX08AI47G), Guest Investigator (NNX09AG31G), MMS theory (NNX08AT76G) and ISIS/Solar Probe Plus theory programs. K.T.O is a member of ISSI team 185.


\begin{thebibliography}{99}

\bibitem{AxfordMcKenzie97}
W.I. Axford \& J.F. McKenzie, in Cosmic Winds and the Heliosphere, ed. J.R. Jokipii, C.P. Sonett, \& M.S. Giampapa, Univ. Arizona Press, 31 (1997).  

\bibitem{HabbalEA95}
S.R. Habbal, R. Esser, M. Guhathakurta, \& R. Fisher, Geophys. Rev. Lett. 22, 1465 (1995).

\bibitem{TuMarsch95} 
C.-Y. Tu \& E. Marsch, Space Sci. Rev. 73, 1 (1995).

\bibitem{BreechEA08}
B. Breech, W.H. Matthaeus, J. Minnie, J.W. Bieber, S. Oughton, C.W. Smith, \& P.A. Isenberg, J. Geophys. Res. 113, A08105 (2008).

\bibitem{OsmanEA11b}
K.T. Osman, M. Wan, W.H. Matthaeus, J.M. Weygand, \& S. Dasso, Phys. Rev. Lett. 107, 165001 (2011).

\bibitem{RichardsonEA95}
J.D. Richardson, K.I. Paularena, A.J. Lazarus, \& J.W. Belcher, Geophys. Rev. Lett. 22, 325 (1995).

\bibitem{MarschTu94}
E. Marsch \& C.Y. Tu, Ann. Geophys. 12, 1127 (1994).

\bibitem{HorburyEA98}
T.S. Horbury, E.A. Lucek, A. Balogh, \& D.J. McComas, Geophys. Res. Lett. 25, 4297, (1998).

\bibitem{MattMont80}
W.H. Matthaeus \& D. Montgomery, Annals of the New York Academy of Sciences, 357, 203 (1980).

\bibitem{CarboneEA90}
V. Carbone, P. Veltri, \& A. Mangeney, Phys. Fluids A, 2, 1487 (1990).

\bibitem{Veltri99}
P. Veltri, Plasma Phys. Contr. Fusion, 41, 787 (1999).

\bibitem{SreenivasanAntonia97}
K.R. Sreenivasan \& R.A. Antonia, Ann. Rev. Fluid Mech. 29, 435 (1997).

\bibitem{Barnes79a}
A. Barnes, in Solar System Plasma Physics, Vol. 1, ed. E.N. Parker, C.F. Kennel, \& L.J. Lanzerotti, Amsterdam:North-Holland, 251 (1979).
  
\bibitem{SahraouiEA09}
F. Sahraoui, M.L. Goldstein, P. Robert, \& Yu. V. Khotyaintsev, Phys. Rev. Lett. 102, 231102 (2009).

\bibitem{HowesEA08}
G.G. Howes, S.C. Cowley, W. Dorland, G.W. Hammett, E. Quataert, \& A.A. Schekochihin, J. Geophys. Res. 113, A05103 (2008). 

\bibitem{OsmanEA11} 
K.T. Osman, W.H. Matthaeus, A. Greco, \& S. Servidio, Astrophys. J. 727, L11 (2011).

\bibitem{Burlaga68} 
L.F. Burlaga, Sol. Phys. 4, 67 (1968).

\bibitem{BorovskyDenton11}
J. E. Borovsky \& M. Denton, Astrophys. J. 739, L61 (2011).

\bibitem{Kol41a}
A.N. Kolmogorov, Dokl. Akad. Nauk SSSR, 30, 301 (1941). [Reprinted in Proc. R. Soc. London, Ser. A 434, 9 (1991)].

\bibitem{Obukhov62b}
A.M. Obukhov, J. Fluid Mech. 13, 77 (1962).

\bibitem{Kol62}
A.N. Kolmogorov, J. Fluid Mech. 12, 82 (1962).

\bibitem{MattGold82a}
W.H. Matthaeus \& M.L. Goldstein, J. Geophys. Res. 87, 6011 (1982).

\bibitem{PolitanoPouquet95}
H. Politano \& A. Pouquet, Phys. Rev. E, 52, 636 (1995).

\bibitem{PolitanoPouquet98-grl}
H. Politano \& A. Pouquet, Geophys. Res. Lett. 25, 273 (1998).

\bibitem{MattLamkin86}
W.H. Matthaeus \& S.L. Lamkin, Phys. Fluids, 29, 2513 (1986).

\bibitem{Smith1998} 
C.W. Smith, M.H. Acuna, L.F. Burlaga, J. L'Heureux, N.F. Ness, \& J. Scheifele, Space Sci. Rev. 86, 611 (1998).
 
\bibitem{McComas1998} 
D.J. McComas, S.J. Blame, P. Barker, W.C. Feldman, J.L. Phillips, P. Riley, \& J.W. Griffee, Space Sci. Rev. 86, 563 (1998).

\bibitem{GrecoEA09}
A. Greco, W.H. Matthaeus, S. Servidio, P. Chuychai, \& P. Dmitruk, Astrophys. J. 691, L111 (2009).

\bibitem{MatthaeusEA05}
W.H. Matthaeus, S. Dasso, J.M. Weygand, L.J. Milano, C.W. Smith, \& M.G. Kivelson, Phys. Rev. Lett. 95, 231101 (2005).

\bibitem{GrecoEA09-wait}
A. Greco, W.H. Matthaeus, S. Servidio, \& P. Dmitruk, Phys. Rev. E, 80, 046401 (2009).

\bibitem{Demoulin09}
P. D\'emoulin, Solar Phys. 257, 169 (2009).

\bibitem{MerrifieldEA05}
J.A. Merrifield, W.C. M\"uller, S.C. Chapman, \& R.O. Dendy, Phys. Plasmas 12, 022301 (2005).

\bibitem{ServidioEA11}
S. Servidio, A. Greco, W.H. Matthaeus, K.T. Osman, \& P. Dmitruk, J. Geophys. Res. 116, A09102 (2011).

\bibitem{CranmerEA09} 
S.R. Cranmer, W.H. Matthaeus, B.A. Breech, and J.C. Kasper, Astrophys. J. 702, 1604 (2009).

\bibitem{VerdiniEA10} 
A. Verdini, M. Velli, W.H. Matthaeus, S. Oughton, \& P. Dmitruk, Astrophys. J. 708, L116 (2010).

\bibitem{LeamonEA99}
R.J. Leamon, C.W. Smith, N.F. Ness, \& H.K. Wong, J. Geophys. Res. 104, 331 (1999).

\end{thebibliography}
\end{document}